\documentclass[sigconf,nonacm]{acmart}

\AtBeginDocument{}
\setcopyright{none}
\acmDOI{}
\acmISBN{}

\usepackage[most]{tcolorbox} % add it
\usepackage{marvosym} % for \Letter symbol
\definecolor{panelBack}{HTML}{F1F4F7}
\definecolor{panelRule}{RGB}{130,136,145}
\newtcolorbox{topmatterBox}{
    enhanced,
    colback=panelBack,
    colframe=panelBack,
    boxrule=0pt,
    frame hidden,
    arc=3mm,
    boxsep=0pt,
    left=6mm, right=6mm, top=5mm, bottom=4mm,
    before skip=4mm, after skip=0mm,
    notitle
}
\newcommand{\headerlogo}{%
  \includegraphics[trim={32.8 43.5 23.4 42.3}, clip,
                   height=4mm]{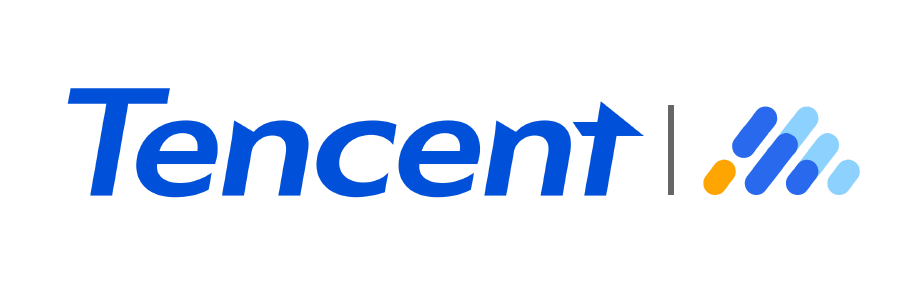}}
\newcommand{\panelnamefont}{\normalsize\bfseries}
\newcommand{\panelaffilfont}{\small}
\newcommand{\panelmailfont}{\ttfamily\footnotesize}
\makeatletter
\newbox\TKF@titlebx
\let\@mktitle\@mktitle@i
\def\@titlefont{\Huge\bfseries}
\def\@mkauthors{\global\setbox\TKF@titlebx=\box\mktitle@bx}
\def\@mkteasers{%
  \global\setbox\mktitle@bx=\vbox{%
    \hsize=\textwidth \linewidth=\textwidth \columnwidth=\textwidth
    \noindent\headerlogo\par
    \vskip 1.25mm % 这里是调整logo到最顶部横线的距离
    {\color{panelRule}\hrule height 1.2pt}%
    \vskip 4.75mm % 这里是调整title到蓝框的距离
    \box\TKF@titlebx
    \begin{topmatterBox}
      \setlength{\parindent}{0pt}% 面版内不进行缩进
      \authorpanel               % 作者 / 单位 / 邮箱
      \medskip
      \noindent\ignorespaces\@abstract\par  % 摘要（取自 \begin{abstract}）
      \ifx\@keywords\@empty\else
        \medskip\noindent{\bfseries\keywordsname:} \@keywords\par
      \fi
      \medskip
      {\color{panelRule}\hrule height 0.4pt}%
      \smallskip
      \noindent{\small\authornotepanel}\par
    \end{topmatterBox}
    \vskip 3.5mm % 面板与双栏正文之间留的空隙
  }%
  \global\let\TKF@concepts\@concepts
  \global\let\TKF@keywords\@keywords
  \global\let\@mkabstract\@empty
  \global\@ACM@printccsfalse
  \global\let\@concepts\@empty
  \global\let\@keywords\@empty}
\let\TKF@printendtopmatter\@printendtopmatter
\def\@printendtopmatter{%
  \global\let\@concepts\TKF@concepts
  \global\let\@keywords\TKF@keywords
  \hypersetup{pdfsubject={\@concepts}, pdfkeywords={\@keywords}}%
  \TKF@printendtopmatter}
\makeatother

\newcommand{\St}{\mathbf{S}}
\newcommand{\NSt}{\mathbf{N}}
\newcommand{\GGt}{\mathbf{G}}
\newcommand{\OGt}{\mathbf{O}}
\newcommand{\numS}{n_{\mathrm{S}}}
\newcommand{\numNS}{n_{\mathrm{N}}}
\newcommand{\numGG}{n_{\mathrm{G}}}
\newcommand{\numOG}{n_{\mathrm{O}}}
\newcommand{\Local}{\operatorname{Local}}
\newcommand{\Fuse}{\operatorname{NSFusion}}
\newcommand{\Global}{\operatorname{GlobalSync}}
\newcommand{\Readout}{\operatorname{CrossAttn}}

\usepackage{multirow}
\usepackage{enumitem}

\title[TokenFormer V2]{TokenFormer V2: Breaking the Resolution--Complexity Dilemma in Unified Recommendation}
\newcommand{\authorpanel}{}
\newcommand{\authornotepanel}{}
\renewcommand{\authorpanel}{%
  {\panelnamefont Kaihui Wu$^{1,*}$, Guoming Li$^{1,2,*,\dagger}$, Liang Guo$^{1,*}$, Yifeng Zhou$^{1,*}$, Yuehong Hu$^{1,*}$, Junwei Pan$^{1}$, Zhixiang Feng$^{1}$, Chengguo Yin$^{1}$, Wenqi Fan$^{2}$, Haijie Gu$^{1}$, Jie Jiang$^{1}$} \\[0.4em]
  {\panelaffilfont $^{1}$Tencent Inc. 
  \quad $^{2}$The Hong Kong Polytechnic University} \\[0.3em]
  {\panelmailfont wukaihui666@gmail.com, guoming.li@connect.polyu.hk, \{mattguo, joefzhou, patrickhu, lionelfeng\}@tencent.com} \\[0.5em]
}
\renewcommand{\authornotepanel}{%
  $^*$Equal contribution. $^\dagger$Work done as part of the Tencent Rhino-Bird Research Program.
}
\renewcommand{\shortauthors}{TokenFormer V2}
\makeatletter
\gdef\authors{}
\makeatother

\begin{document}
\begin{abstract}
  \textbf{Abstract.}
Industrial recommendation faces a resolution--complexity dilemma: dense interactions make detailed modeling of long behavior histories expensive, while compact history summaries restrict the evidence available to prediction. A unified backbone must preserve rich behavioral details within strict serving budgets. We introduce \textbf{TokenFormer V2}, a unified architecture that explicitly decouples \emph{representation resolution} from \emph{interaction density}. Its group-wise token layout jointly encodes behavior segments with non-sequential context, expands interaction ranges via hierarchical group merging, and coordinates distributed groups through \emph{NS Fusion} and \emph{Global Sync} modules. Through this \emph{local--global decomposition} paradigm, computation is concentrated on structured interactions while preserving fine-grained evidence, ensuring that every encoded event remains directly accessible to the \emph{Full-Resolution Readout} module. On the Tencent Advertising Algorithm Competition 2026 (TAAC 2026) benchmark, TokenFormer V2 achieves top-ranking accuracy among unified recommendation baselines while using \textbf{5.1$\times$ fewer FLOPs} than leading baselines, sustaining its advantage across long histories and imbalanced behavior domains. Furthermore, in billion-parameter evaluations on WeChat Channels advertising, it consistently outperforms state-of-the-art industrial baselines, yielding up to a \textbf{+0.07 pp} overall AUC gain. In live online A/B testing against the production model, TokenFormer V2 delivers a statistically significant \textbf{2.33\% increase in GMV}. These results establish TokenFormer V2 as a battle-tested blueprint for scalable, full-resolution behavior modeling in industrial recommendation.
\end{abstract}
\maketitle
\begin{figure*}[t]
  \centering
  \includegraphics[width=\textwidth]{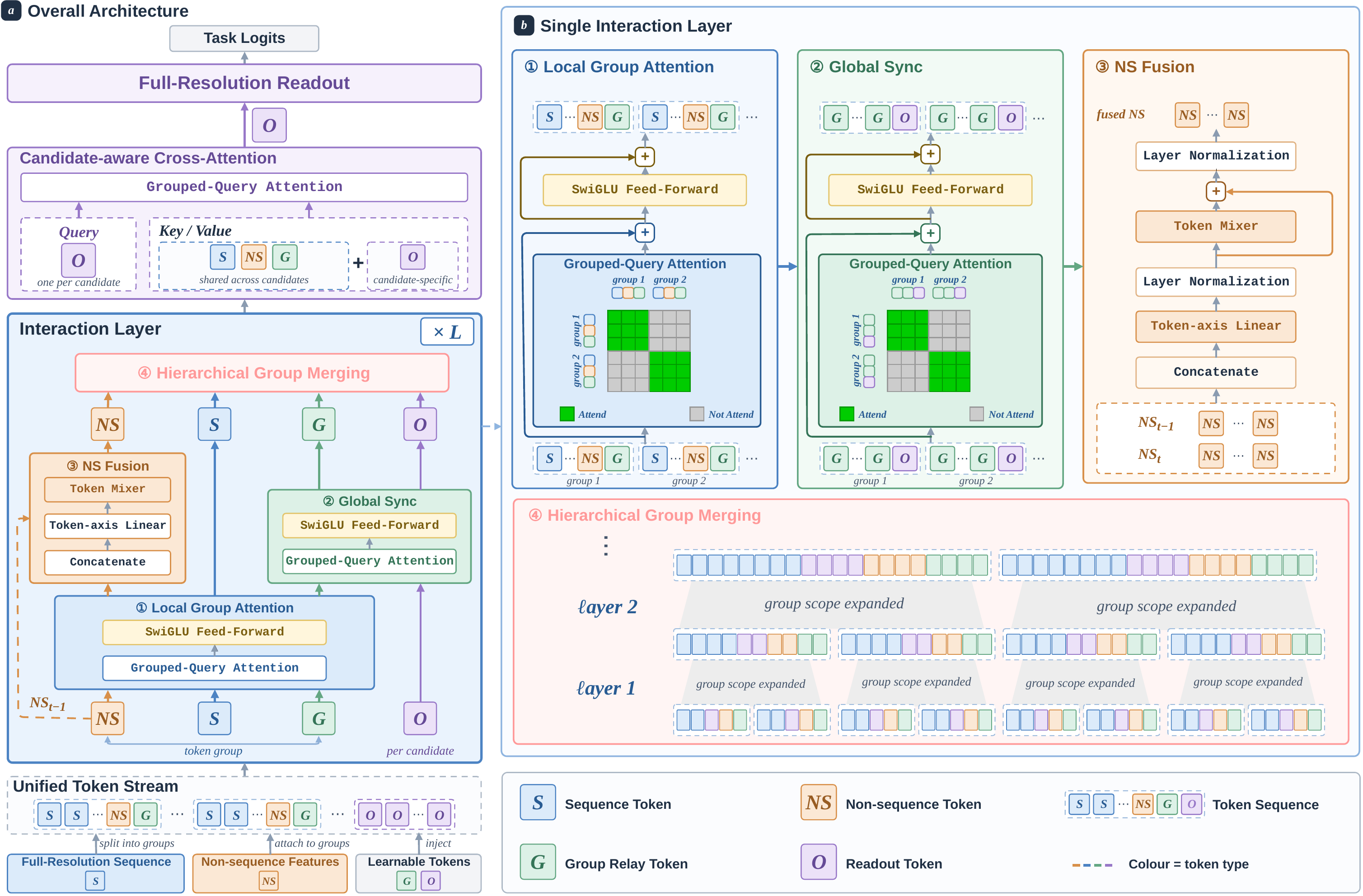}
  \caption{Overview of TokenFormer V2, which separates representation resolution from interaction density. (a) Full-resolution sequence passes through hierarchical interaction layers and remains directly accessible to the candidate-aware Full-Resolution Readout. (b) Local Group Attention, NS Fusion, and Global Sync coordinate local evidence and shared context; Hierarchical Group Merging expands group scope without compression, preserving individual representations throughout the backbone.}
  \Description{The supplied architecture diagram has two panels. Panel a traces the Full-Resolution Sequence through repeated interaction layers, candidate-aware cross-attention, and Full-Resolution Readout to task logits. Panel b illustrates Local Group Attention, Global Sync, NS Fusion, and Hierarchical Group Merging with expanding group scope. Behavior, non-sequential, group relay, and readout tokens are distinguished by color.}
  \label{fig:layout}
\end{figure*}

% Keep paragraph spacing natural in this partial manuscript.
\raggedbottom

% Explain token retention and grouped interactions before naming the modules.
% Token names: group relay tokens (G) and readout tokens (O).

\section{Introduction}
\label{sec:introduction}

Industrial ranking models combine user behavior histories with non-sequential features describing the user, candidate item, and request context. Research on these inputs has historically developed along two complementary branches: sequence modeling extracts user interests from past actions~\cite{zhou2018din,pi2020sim,chang2023twin}, while feature-interaction models learn crosses among non-sequential fields~\cite{wang2021dcnv2}. Conventional systems often encode the history first and then feed its summary into a feature-interaction module. Recent unified recommendation Transformers bring these two branches into a shared backbone, allowing sequential and non-sequential representations to interact seamlessly throughout the model~\cite{zhang2025onetrans,huang2026hyformer,huang2026mixformer}.

Existing unified models explore various ways to organize this interaction. OneTrans places behavior and non-sequential tokens in a shared causal stream~\cite{zhang2025onetrans}. HyFormer and MixFormer alternate history aggregation through feature-derived queries with query or feature mixing~\cite{huang2026hyformer,huang2026mixformer}. EST and Kunlun combine feature interaction with sparse behavior propagation and hierarchical sequence summarization, respectively~\cite{liu2026est,hou2026kunlun}. Our earlier \textbf{TokenFormer}~\cite{zhou2026tokenformer} unifies field, behavior, and target tokens, using Bottom-Full-Top-Sliding attention and nonlinear multiplicative interactions to improve representation learning. Together, these methods have significantly advanced the design of interaction operators and the information flow between sequential and non-sequential inputs.

Long histories, however, make it difficult to give every behavior rich contextual modeling within a practical compute budget. For example, several nearby actions may together indicate an intent that is unclear from any single action. Dense self-attention can model these relationships, but incurs a quadratic cost in sequence length~\cite{vaswani2017attention}; TokenFormer's shallow full-attention layers retain this cost even though its upper layers use narrower windows. To reduce computation, OneTrans progressively shrinks the set of behavior queries, leaving fewer event representations for deeper layers. HyFormer and MixFormer concentrate joint field--sequence interaction in compact query states, so subsequent feature interactions operate on aggregated history representations. These strategies are efficient, but leave a vital design goal unmet: \emph{contextualize every retained behavior with surrounding events and field information, keep distant history segments connected, and preserve direct access to individual events for prediction.} Meeting this goal requires deciding precisely where detailed interactions are needed and how information travels between them.

Building on TokenFormer's unified modeling, we introduce \textbf{TokenFormer V2}. Instead of compressing history into a compact summary, our approach is built on a direct premise: we preserve an individual token for every retained behavior throughout encoding. Each behavior interacts locally with neighboring events and field context within its group, while a small set of communication tokens bridges the groups. This ensures that the final prediction can still attend directly to every encoded event—a property we term \emph{full resolution}. Our \emph{token layout} serves as the organizing principle that coordinates these local groups and their communication paths.

Specifically, we partition the behavior sequence into contiguous segments and assign each segment a subset of non-sequential (NS) tokens and learned \emph{group relay tokens}. The \emph{Local Group Attention} updates the tokens within each group, so each behavior representation incorporates neighboring events and field context. As depth increases, \emph{Hierarchical Group Merging} combines adjacent groups. For instance, merging two groups of four behavior tokens creates a group with eight behavior tokens, each enjoying a wider interaction scope. Merging therefore broadens direct interaction without ever pooling or discarding individual behaviors.

Meanwhile, two compact communication paths connect these local computations at every layer. \emph{NS Fusion} mixes field representations updated by different local histories and redistributes the resulting context to the groups. \emph{Global Sync} exchanges information among the group relay tokens and a small set of learned \emph{readout tokens}. The relay tokens carry this shared information back into local encoding, while the readout tokens accumulate information for prediction. Finally, \emph{Full-Resolution Readout} uses the readout tokens as queries for candidate-aware cross-attention over the complete encoded token set. The model thus preserves individual behavior tokens for prediction while strictly controlling attention cost through group sizes and compact communication paths.

We evaluate TokenFormer V2 across both large-scale benchmarks and a production advertising system. On the Tencent Advertising Algorithm Competition 2026 (TAAC 2026) benchmark, it achieves the highest AUC and lowest log loss among eight evaluated backbones, using only $19.5\%$ of the forward FLOPs of MixFormer, the strongest predictive baseline. Its advantage persists when the history length doubles and across user segments featuring imbalanced behavior domains. Furthermore, on historical WeChat Channels advertising logs, a billion-parameter model improves overall AUC over MixFormer by \textbf{0.0572 pp} with identical input features; a variant with simplified inputs further increases the gain to \textbf{0.0712 pp}. In live online A/B testing against the production baseline, it delivers a statistically significant \textbf{2.33\% increase in GMV}.

Our contributions are threefold:

\begin{itemize}[leftmargin=15pt]
  \item \textbf{A group-wise token layout for long histories.} We combine field-aware local groups with hierarchical merging to expand contextual interactions while preserving an individual token for every retained behavior.
  \item \textbf{Cross-group communication with full-resolution readout.} NS Fusion and Global Sync connect local groups through compact context tokens, while the final readout retains direct access to every encoded behavior.
  \item \textbf{Validation from benchmark to production.} We demonstrate accuracy and efficiency gains over strong unified baselines, examine long and imbalanced histories, and validate billion-parameter modeling and statistically significant online GMV gains on WeChat Channels advertising.
\end{itemize}

\section{Related Work}
\label{sec:related}

\paragraph{\bf Feature interaction.}
Feature interaction models capture dependencies among user, item, and context fields. DCN-V2~\cite{wang2021dcnv2} learns explicit feature crosses with matrix-parameterized cross layers and mixtures of low-rank experts. RankMixer~\cite{zhu2025rankmixer} combines parameter-free token mixing with per-token feed-forward transformations. TokenFormer V2 incorporates field representations throughout behavior encoding, so contextual features help shape the historical evidence consumed by prediction.

\paragraph{\bf Behavior modeling.}
Behavior models learn which past actions inform a recommendation. DIN~\cite{zhou2018din} uses candidate-aware aggregation to extract relevant interests, while SASRec~\cite{kang2018sasrec} contextualizes sequential behavior with causal self-attention. HSTU~\cite{zhai2024hstu} formulates recommendation as sequential transduction over heterogeneous user actions. These lines of work motivate two complementary capabilities: interpreting events in their historical context and selecting the evidence relevant to the current prediction.

\paragraph{\bf Unified recommenders.}
Recent architectures integrate feature interaction and sequence modeling, with token organization playing a central role. OneTrans~\cite{zhang2025onetrans} combines non-sequential and behavior tokens in a causal stream and progressively reduces behavior queries through a pyramidal schedule. TokenFormer~\cite{zhou2026tokenformer}, an earlier unified backbone, organizes field, behavior, and target tokens in a causal stream. Its Bottom-Full-Top-Sliding attention uses full causal attention in shallow layers and narrower windows at greater depth. HyFormer~\cite{huang2026hyformer} and MixFormer~\cite{huang2026mixformer} use feature-derived queries to extract compact history representations, coupled with feature interaction. EST~\cite{liu2026est} combines lightweight field--behavior cross-attention with sparse propagation among similar behaviors; Kunlun~\cite{hou2026kunlun} couples hierarchical sequence summarization with global feature interaction. Building on TokenFormer's token-level unification, TokenFormer V2 develops a group-wise organization of the backbone: field context participates inside local behavior groups, the groups merge with depth, and compact global paths coordinate a persistent set of individual behavior representations. Its contribution is the resulting local--global architecture and its demonstrated accuracy--efficiency advantage from benchmark evaluation to live production.

% File prefixes reproduce the requested project organization, not section IDs.
\section{TokenFormer V2}
\label{sec:model}
TokenFormer V2 separates three responsibilities: \emph{local interpretation} of behavior in field context, \emph{global coordination} across the history, and \emph{candidate selection} of the resulting evidence. Figure~\ref{fig:layout} connects them through a full-resolution sequence, hierarchical interaction layers, and Full-Resolution Readout. Hierarchical Group Merging expands each group's interaction scope with depth, while NS Fusion and Global Sync exchange context at every layer. The token inventory remains fixed throughout the hierarchy, and the final prediction stage retains direct access to every encoded behavior.

\subsection{Inputs and Token Representations}
\label{sec:tokens}

For each prediction instance, we represent the inputs by four types of $d$-dimensional tokens: behavior tokens $\St$, non-sequential tokens $\NSt$, group relay tokens $\GGt$, and readout tokens $\OGt$. Table~\ref{tab:notation} gives their counts. Each representation matrix has one row per token. The model has $L$ layers, and brackets $[A;B]$ denote concatenation along the token axis. Figure~\ref{fig:layout} labels $\NSt$, $\GGt$, and $\OGt$ as NS, G, and O, respectively.

\begin{table}[!t]
  \caption{Token types and counts. Each token has width $d$.}
  \label{tab:notation}
  \centering
  \begin{tabular}{@{}lll@{}}
    \toprule
    Symbol & Token type & Count \\
    \midrule
    $\St$   & Behavior tokens & $\numS$ \\
    $\NSt$  & Non-sequential tokens & $\numNS$ \\
    $\GGt$  & Group relay tokens & $\numGG$ \\
    $\OGt$  & Readout tokens & $\numOG$ \\
    \bottomrule
  \end{tabular}
\end{table}

\paragraph{\bf Full-Resolution Sequence.}
The Full-Resolution Sequence $\St^{(0)}$ contains $\numS$ valid behavior tokens, one for each retained event. These individual representations persist through all interaction layers. Token $j$ has layout position $p_j\in\{0,\ldots,T-1\}$, where $T$ is the maximum history length. This index determines group membership and is retained when padding is removed. For multi-domain histories, each token also carries its within-domain position $r_j$ for positional encoding; the two indices serve distinct roles.

\paragraph{\bf Non-sequential tokens.}
User, candidate, and request fields are encoded together into one set of non-sequential tokens. Let $\mathbf X\in\mathbb{R}^{f\times d}$ contain the embeddings of the $f$ fields. We construct
\begin{equation}
  \NSt^{(0)}=\operatorname{Entry}(\mathbf X)
  \in\mathbb{R}^{\numNS\times d}.
  \label{eq:entry}
\end{equation}
For each token, we concatenate the corresponding channel slice from every field, apply a shared projection, and add a token-specific feed-forward residual. Each non-sequential token therefore receives input from every field. Appendix~\ref{app:entry} gives the full formulation.

\paragraph{\bf Relay and readout tokens.}
$\GGt^{(0)}$ and $\OGt^{(0)}$ start from learned vectors. Group relay tokens connect local behavior encoding to history-wide communication. Readout tokens accumulate information across layers and provide the queries for Full-Resolution Readout; in the TAAC instantiation, their initialization additionally receives a candidate-conditioned residual (Appendix~\ref{app:field_tokens}). The counts $\numS$, $\numNS$, $\numGG$, and $\numOG$ remain fixed across layers.

\subsection{Group-wise Token Layout}
\label{sec:layout}

The layout controls connectivity while preserving the token inventory. Local groups provide dense interaction where behaviors are interpreted together; the hierarchy broadens that context without replacing individual events by group summaries.

\paragraph{\bf Assigning tokens to groups.}
At layer $\ell$, we partition the original history positions into $b_\ell$ equal, contiguous intervals. A behavior at position $p_j$ belongs to group $1+\lfloor p_jb_\ell/T\rfloor$. We also partition $\NSt$ and $\GGt$ into consecutive, disjoint subsets, assigning $\numNS/b_\ell$ non-sequential tokens and $\numGG/b_\ell$ group relay tokens to each group. Each token in $\St$, $\NSt$, and $\GGt$ belongs to one group per layer. The group counts divide $T$, $\numNS$, and $\numGG$.

Within a single layer, we write $b=b_\ell$ and omit the layer index. Unmarked representations denote the layer input, an overline denotes the local update, and a superscript $+$ denotes the layer output. For group $i\in\{1,\ldots,b\}$, the token arrangement and local update are
\begin{equation}
  \begin{aligned}
    \mathbf A_i&=[\St_i;\NSt_i;\GGt_i],\\
    \overline{\mathbf A}_i&=\Local(\mathbf A_i).
  \end{aligned}
  \label{eq:layout}
\end{equation}
Within each group containing valid behaviors, all tokens attend to one another. Field context therefore participates in the formation of each behavior representation, and the same update enriches the context tokens with evidence from that segment. NS Fusion and Global Sync then communicate across groups (Section~\ref{sec:architecture}).

\paragraph{\bf Hierarchical Group Merging.}
The partitions form a nested hierarchy across layers. Hierarchical Group Merging combines adjacent groups to expand their direct interaction scope, as illustrated in Figure~\ref{fig:layout}(b). For example, $b_{\ell+1}=b_\ell/2$ combines two neighboring behavior intervals and their assigned non-sequential and group relay tokens. Each merged group retains an individual representation for every constituent token.

The expanded scope gives each token a larger contextual neighborhood without pooling or discarding tokens. Early layers can resolve segment-level patterns, while deeper layers combine these patterns over larger portions of the history. Cross-group communication is already available at every depth, so global context does not have to wait for the hierarchy to reach a single group. The final layer can contain one or more groups.

\paragraph{\bf Multi-domain histories.}
For multi-domain histories, we concatenate the domains into a single sequence, ordering each domain from its newest event backwards. Group membership uses the resulting layout position $p_j$; RoPE uses the within-domain position $r_j$, and a learned domain embedding identifies the source of each event. Groups may therefore span a domain boundary while retaining both temporal and domain identity. Non-sequential, relay, and readout tokens connect the domains throughout the backbone, and every available behavior retains its own route to Full-Resolution Readout. For production serving, the history can also be split into branches with different depths, as described in Section~\ref{sec:exp_online}.

\subsection{Local-Global Interaction Enabling Full-Resolution Readout}
\label{sec:architecture}

\paragraph{\bf Local Group Attention.}
The $\Local$ block in Eq.~\eqref{eq:layout} applies normalized, gated self-attention and a feed-forward network, each with a residual connection. Attention is bidirectional within each group, with shared attention parameters across groups in the same layer. Rotary position embeddings (RoPE)~\cite{su2021roformer} are applied to behavior queries and keys using within-domain positions $r_j$. Non-sequential, relay, and readout tokens use unrotated queries and keys.

Reassembling the local outputs in the original token order gives $\overline{\St}$, $\overline{\NSt}$, and $\overline{\GGt}$. Groups without valid behaviors carry their input $\NSt$ and $\GGt$ representations to the cross-group updates. As shown in Figure~\ref{fig:layout}(b), the complete interaction layer is
\begin{equation}
  \begin{aligned}
    \St^+&=\overline{\St},\\
    \NSt^+&=\Fuse(\NSt,\overline{\NSt}),\\
    [\GGt^+;\OGt^+]&=\Global([\overline{\GGt};\OGt]).
  \end{aligned}
  \label{eq:layer}
\end{equation}
The behavior tokens retain the detailed outputs of local encoding. Non-sequential, relay, and readout tokens provide two complementary communication paths: one refreshes shared field context, and the other coordinates history-wide information. The readout tokens enter each layer through Global Sync.

\paragraph{\bf NS Fusion.}
NS Fusion turns local field--behavior interactions into context for the next layer. It combines the incoming non-sequential representations with their locally updated counterparts, so both the existing field context and newly encoded evidence contribute to the update. A linear transformation along the token axis mixes the concatenated representations into $\numNS$ tokens, followed by a residual token mixer based on RankMixer~\cite{zhu2025rankmixer}. Redistributing these tokens to the next layer lets each group receive context enriched by the other segments. Appendix~\ref{app:fusion} specifies the operators.

\paragraph{\bf Global Sync.}
Global Sync gives every local group access to a shared global communication space. It applies self-attention to $[\overline{\GGt};\OGt]$, followed by a feed-forward network, with normalization, attention-output gating, and residual connections. These relay and readout tokens attend to one another: $\GGt^+$ carries the resulting context back into local encoding, while $\OGt^+$ develops the queries used for prediction. The compact global path coordinates the groups while their behavior tokens remain distinct. Each layer has its own local, fusion, and global parameters.

\paragraph{\bf Full-Resolution Readout.}
Full-Resolution Readout is the final prediction stage, combining candidate-aware cross-attention with a feed-forward transformation and task-specific output mappings. Its full resolution comes from direct access to every retained behavior after contextual encoding. After $L$ interaction layers, the attention step uses $\OGt^{(L)}$ as queries and the complete encoded token set as keys and values:
\begin{equation}
  \begin{aligned}
    \mathbf Z&=\Readout\!\left(\OGt^{(L)};
    [\St^{(L)};\NSt^{(L)};\GGt^{(L)};\OGt^{(L)}]\right),\\
    \widehat y_k&=\operatorname{Head}_k\!\left(
    \operatorname{vec}(\operatorname{FFN}(\mathbf Z))\right).
  \end{aligned}
  \label{eq:readout}
\end{equation}
Here $k$ indexes prediction tasks. Cross-attention uses normalization, output gating, and a residual connection to its queries. RoPE is applied to behavior keys using $r_j$; other keys and the readout queries remain unrotated. The output FFN uses token-specific residual transformations followed by RMS normalization. The task-specific mappings produce logits and binary response probabilities, trained with binary cross-entropy.
\subsection{Computational Complexity}
\label{sec:complexity}

TokenFormer V2 preserves fine-grained behavior tokens while strictly controlling interaction density through its group schedule. 
At layer $\ell$, the layout partitions $N=\numS+\numNS+\numGG$ behavior and context tokens into $b_\ell$ disjoint groups. Let $n_{\mathrm{S},i}^{(\ell)}$ denote the number of valid behavior tokens in group $i$. Each group also receives $(\numNS+\numGG)/b_\ell$ context tokens. For hidden width $d$, the local-attention cost is bounded by
\begin{equation}
  \mathcal O\!\left(
    d\sum_{i=1}^{b_\ell}
    \left[
      n_{\mathrm{S},i}^{(\ell)}
      +\frac{\numNS+\numGG}{b_\ell}
    \right]^2
  \right).
  \label{eq:cost}
\end{equation}
Let $m_\ell$ be the maximum group size including both behavior and context tokens. Since every token belongs to exactly one group, Eq.~\eqref{eq:cost} is at most $\mathcal O(d\,m_\ell N)$, which is linear in the total token count $N$ for a bounded group size. For balanced groups, it becomes $\mathcal O(dN^2/b_\ell)$, explicitly exposing $b_\ell$ as a knob for interaction cost. Across $L$ layers, the local-attention complexity totals $\mathcal O(dN\sum_{\ell=1}^{L}m_\ell)$.

The remaining modules add minimal overhead: NS Fusion and Global Sync depend solely on compact context-token counts, while Full-Resolution Readout adds $\mathcal O(d\,\numOG(N+\numOG))$ attention cost via $\numOG$ queries, maintaining linearity in history length. Projections and feed-forward layers are also linear in $N$. Detailed term-by-term derivations are provided in Appendix~\ref{app:cost}.
\begin{table*}[!t]
  \caption{Accuracy and efficiency on TAAC 2026 at approximately 11M dense parameters. Best predictive metrics are bold; tied second-best values are underlined.}
  \label{tab:main}
  \centering
  \begin{tabular}{@{}lcccccc@{}}
    \toprule
    Model & Test AUC $\uparrow$ & Log loss $\downarrow$ & GFLOPs & Train (samples/s) & Inference ($\mu$s/sample) & Dense params \\
    \midrule
    Transformer~\cite{vaswani2017attention} & 0.8168 & 0.2288 & 32.59 & 161 & 2921 & \multirow{8}{*}{$\sim$11M} \\
    TokenFormer~\cite{zhou2026tokenformer} & 0.8198 & 0.2281 & 28.52 & 333 & 1945 & \\
    OneTrans~\cite{zhang2025onetrans} & 0.8192 & 0.2281 & 1.90 & 923 & 157 & \\
    HyFormer~\cite{huang2026hyformer} & 0.8204 & \underline{0.2270} & 3.13 & 900 & 131 & \\
    EST~\cite{liu2026est} & 0.8159 & 0.2290 & 2.75 & 779 & 271 & \\
    Kunlun~\cite{hou2026kunlun} & 0.8176 & 0.2288 & 2.05 & 929 & 171 & \\
    MixFormer~\cite{huang2026mixformer} & \underline{0.8211} & \underline{0.2270} & 19.75 & 537 & 727 & \\
    \cmidrule(r){1-6}
    \textbf{TokenFormer V2} & \textbf{0.8221} & \textbf{0.2261} & 3.86 & 651 & 282 & \\
    \bottomrule
  \end{tabular}
\end{table*}

\section{Offline Experiments}
\label{sec:experiments}

Our offline evaluation tests the central promise of the group-wise layout: stronger prediction from full-resolution histories at a practical computational cost. We establish the accuracy--efficiency advantage on TAAC 2026, examine its behavior as histories grow and domain coverage varies, analyze the predictive information exposed by its representations, and isolate the roles of local encoding and global communication. Section~\ref{sec:exp_online} then evaluates the architecture in the WeChat Channels production setting.

\subsection{Evaluation Protocol}
\label{sec:exp_setup}

\paragraph{\bf Data and prediction task.}
The TAAC 2026 benchmark was released for the \emph{TAAC $\times$ KDD Cup 2026 Tencent Uni-Rec Challenge}.\footnote{Official competition website: \url{https://algo.qq.com}.} It provides 47.1 million impressions with conversion labels, user and candidate fields, dense features, and histories from four behavior domains, denoted A--D. We retain the 256, 256, 512, and 512 most recent events from these domains, respectively, for a total of up to 1{,}536 behaviors per impression. Domain coverage is highly uneven: the median history in domain D is more than three times that in any other domain. Of the 34.8 million training impressions, the last 30 files (0.35 million impressions) form the validation set. The test set contains 12.3 million later impressions from users absent from training. This chronological, user-disjoint split tests prediction on a subsequent population rather than repeated impressions of known training users. Further data details appear in Appendix~\ref{app:offline_diagnostics}.

\paragraph{\bf Models and training.}
Table~\ref{tab:main} compares TokenFormer V2 with seven unified backbones spanning full token streams, hierarchical sequence processing, and feature-query aggregation. Each baseline retains its native architecture. All models share the input features, data pipeline, 64-dimensional embedding tables (623.5M parameters), and prediction head, with approximately 11M dense parameters per model. We train for one epoch at batch size 1{,}024 using AdamW for dense parameters, Adagrad for embeddings, and bf16 precision. Architectural and optimization hyperparameters are selected on validation data; Appendix~\ref{app:exp_details} provides the training recipe and selected configurations.

\paragraph{\bf TokenFormer V2 configuration.}
The four domains form one Full-Resolution Sequence with domain embeddings and within-domain positions. Three interaction layers of width $d=192$ use the group schedule $16\rightarrow8\rightarrow4$, expanding the local behavior span from 96 to 192 to 384 positions. The backbone has 64 non-sequential, 32 group relay, and 4 readout tokens, with grouped-query attention using 6 query heads and 2 key--value heads. Field tokenization is specified in Appendix~\ref{app:field_tokens}.

\paragraph{\bf Metrics and compute.}
We report test AUC and log loss, forward GFLOPs per sample, training throughput per NVIDIA H20 GPU, and amortized inference latency per sample. Inference uses batch size 1{,}024, or 256 for the standard Transformer. AUC differences are reported in percentage points (pp), so an increase of 0.001 in AUC is 0.1~pp.

\subsection{Stronger Prediction per Unit of Compute}
\label{sec:exp_main}

TokenFormer V2 achieves the highest AUC and lowest log loss among all eight models in Table~\ref{tab:main}. Against MixFormer, the strongest baseline, it raises AUC by 0.10~pp and reduces log loss by 0.40\%, using just 19.5\% of its forward FLOPs. The advantage therefore combines better ranking with a substantial reduction in the computation needed to obtain it. Inference is $2.6\times$ faster and training throughput is 21\% higher under the reported batch settings.

The comparison with full-resolution token streams highlights the value of organizing interactions. Relative to TokenFormer, TokenFormer V2 improves AUC by 0.23~pp while reducing FLOPs by $7.4\times$ and amortized inference latency by $6.9\times$. Against the standard Transformer, the AUC gain reaches 0.53~pp at $8.4\times$ fewer FLOPs. These gains come from a backbone that retains an individual representation for every behavior throughout encoding. The group hierarchy concentrates direct interaction within coherent history segments, while the context, relay, and readout tokens keep those segments connected.

Efficient query-based architectures provide a complementary comparison. TokenFormer V2 surpasses HyFormer by 0.17~pp at a forward cost of 3.86 versus 3.13 GFLOPs, placing full-resolution encoding in a comparable compute range while delivering the best predictive quality. Together, the results establish a strong operating point for unified recommendation: leading accuracy, explicit access to every historical event, and a backbone cost suitable for large-scale ranking.

\begin{figure}[!t]
  \centering
  \includegraphics[width=\linewidth]{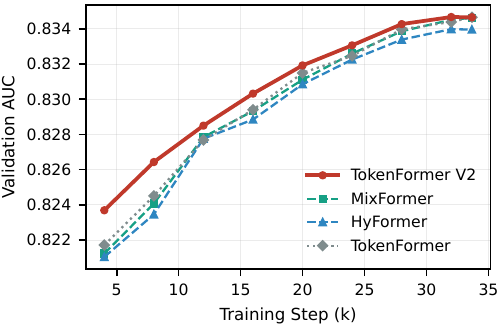}
  \caption{Validation AUC over one training epoch. TokenFormer V2 establishes its lead early and remains ahead at every measured checkpoint.}
  \Description{Validation AUC against training steps for TokenFormer V2, MixFormer, HyFormer and TokenFormer. TokenFormer V2 has the highest curve at every measured checkpoint.}
  \label{fig:val_curves}
\end{figure}

\paragraph{\bf An advantage established early in training.}
\label{sec:learning_dynamics}
Figure~\ref{fig:val_curves} traces validation AUC over the shared training schedule, with evaluation every 4k steps during one epoch. TokenFormer V2 leads at every measured checkpoint. Its advantage is already established at 4k steps, when its AUC exceeds TokenFormer, HyFormer, and MixFormer by 0.20, 0.27, and 0.25~pp, respectively, and persists through the end of the 33.7k-step run. Thus the final ranking advantage is accompanied by more effective learning throughout the training budget. Read together with Table~\ref{tab:main}, the curves connect two useful properties of the layout: stronger prediction at a given training stage and a lower computational cost than the strongest baseline. The same lead then carries into the later, user-disjoint test population.

\subsection{Full-Resolution Histories at Scale}
\label{sec:exp_eff}

The practical test of retaining every behavior is whether the model can continue to do so as histories grow. We scale all four domain lengths by $\{0.25,0.5,1,2\}$, giving 384 to 3{,}072 behaviors, and measure training time and peak memory with fully populated histories at batch size 256 on one H20 GPU. This profile measures the cost of processing the entire configured history.

\begin{figure}[t]
  \centering
  \includegraphics[width=\linewidth]{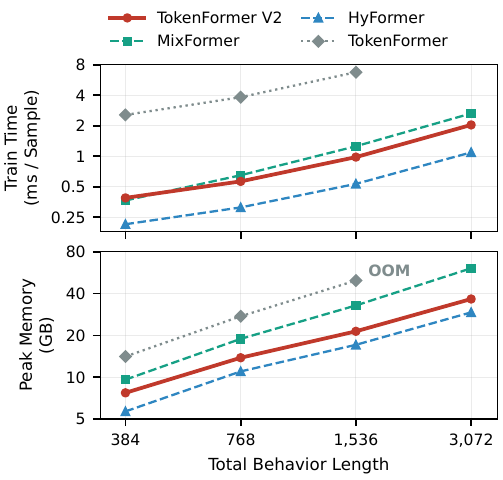}
  \caption{Training time and peak memory as fully populated histories grow from 384 to 3{,}072 events, with batch size 256.}
  \Description{Two log-log plots of training time and peak memory against history length. TokenFormer V2 follows growth close to HyFormer and below MixFormer. TokenFormer requires substantially more time and exceeds device memory at the largest history.}
  \label{fig:scaling}
\end{figure}

Figure~\ref{fig:scaling} shows that an $8\times$ increase in history length raises TokenFormer V2's measured training time by $5.3\times$ and memory by $4.7\times$. Its growth closely tracks HyFormer ($5.1\times$ for both measures), even though TokenFormer V2 retains all behavior representations. At 3{,}072 events, TokenFormer V2 trains 23\% faster than MixFormer with 40\% less memory. TokenFormer requires $6.6$--$6.9\times$ as much training time over the lengths it can process and exceeds the 96GB device memory at the largest length. The group-wise layout thus brings the measured scaling behavior of full-resolution encoding into the range of query-based models over this history-length range.

\subsection{Accuracy Beyond Default History Budget}
\label{sec:exp_long}

We also retrain the four models with twice the history: 512, 512, 1{,}024, and 1{,}024 events from domains A--D, for 3{,}072 events in total. The remaining training recipe is unchanged, including gradient accumulation when needed (Appendix~\ref{app:exp_details}); the fixed-batch memory profile above is a separate efficiency measurement.

TokenFormer V2 leads at both history lengths (Table~\ref{tab:long}). At 3{,}072 events, it exceeds MixFormer by 0.09~pp and TokenFormer by 0.18~pp while requiring $4.4\times$ and $7.6\times$ fewer FLOPs, respectively. Its default 1{,}536-event configuration also outperforms every comparison model given twice as much history. This is a useful deployment property: the architecture extracts enough predictive value at the default budget to remain ahead of competitors with a larger input budget, while preserving its lead when more history is available.

\begin{table}[!t]
  \caption{Accuracy and forward cost after doubling the history. TokenFormer V2 leads at both input budgets.}
  \label{tab:long}
  \centering
  \setlength{\tabcolsep}{3pt}
  \begin{tabular}{@{}lcccc@{}}
    \toprule
    & \multicolumn{2}{c}{1{,}536 events} & \multicolumn{2}{c}{3{,}072 events} \\
    \cmidrule(lr){2-3}\cmidrule(l){4-5}
    Model & AUC & GFLOPs & AUC & GFLOPs \\
    \midrule
    TokenFormer & 0.8198 & 28.52 & 0.8201 & 68.94 \\
    HyFormer & 0.8204 & 3.13 & 0.8204 & 6.22 \\
    MixFormer & 0.8211 & 19.75 & 0.8210 & 39.50 \\
    \textbf{TokenFormer V2} & \textbf{0.8221} & 3.86 & \textbf{0.8219} & 9.07 \\
    \bottomrule
  \end{tabular}
\end{table}

\subsection{Robustness to Uneven Behavioral Evidence}
\label{sec:exp_where}

A unified recommender must serve users whose interests are expressed through very different combinations of domains. We partition test users by their \emph{dominant domain}, which supplies the largest share of their events, and by \emph{inactive domains}, in which they have no events. Domain D dominates approximately three quarters of test impressions, making users dominated by A--C an informative test of coverage beyond the majority population.

\begin{figure}[t]
  \centering
  \includegraphics[width=\linewidth]{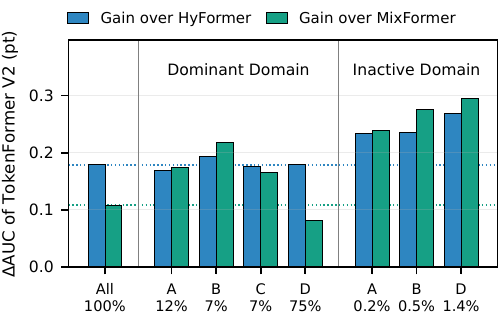}
  \caption{AUC gains across dominant-domain and inactive-domain user segments. TokenFormer V2 improves every reported segment, with larger gains away from the dominant-domain majority.}
  \Description{Grouped bars of AUC gains over HyFormer and MixFormer. All gains are positive. Gains over MixFormer are about 0.08 percentage points for users dominated by D, 0.17 to 0.22 for users dominated by A to C, and up to 0.30 for users with an inactive domain.}
  \label{fig:groups}
\end{figure}

TokenFormer V2 improves over both HyFormer and MixFormer in every reported segment (Figure~\ref{fig:groups}). For users dominated by domains A--C, its gain over MixFormer is 0.17--0.22~pp, compared with 0.08~pp for users dominated by D. For users with an inactive domain, the gain reaches 0.24--0.27~pp over HyFormer and 0.24--0.30~pp over MixFormer. The advantage is therefore particularly pronounced when the available behavioral evidence departs from the dominant-domain pattern.

This pattern accords with the layout's treatment of evidence. Each observed behavior keeps its own contextual representation, while shared non-sequential, relay, and readout tokens connect information across groups. Users with sparse or absent domains can draw on the events they do have, and Full-Resolution Readout retains direct access to those events. The segment results show that the overall accuracy gain extends to heterogeneous users rather than relying on the largest domain alone.

\subsection{Richer Behavioral Representations via Full-Resolution Access}
\label{sec:exp_repr}

Preserving individual behavior tokens is valuable when their evidence becomes useful to the final prediction. We examine this connection through the common 128-dimensional pre-prediction representation produced by each backbone. TokenFormer V2, TokenFormer, HyFormer, and MixFormer are evaluated on the same 200k test impressions. For each model, a logistic-regression probe is fitted on one half of the users and evaluated on the other, providing a common, simple decoder of conversion-related information. This comparison complements the task metrics: it asks how much predictive signal is readily accessible in the representation supplied to the prediction head.

\begin{figure}[!t]
  \centering
  \includegraphics[width=\linewidth]{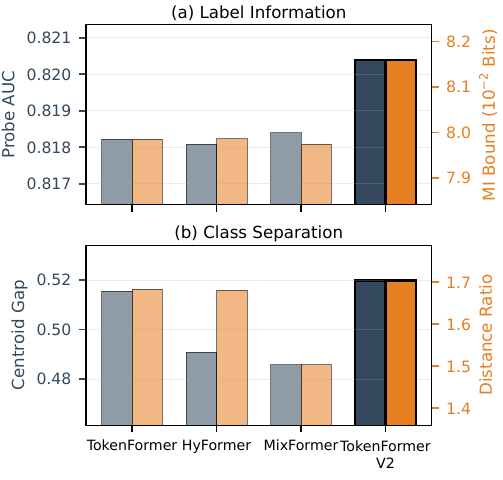}
  \caption{Predictive information in the common 128-dimensional representation. TokenFormer V2 leads in probe AUC and the MI lower bound (a), and in normalized class separation (b).}
  \Description{Two dual-axis bar charts compare TokenFormer, HyFormer, MixFormer and TokenFormer V2. TokenFormer V2 has the highest probe AUC, mutual-information lower bound, class-centroid gap, and distance ratio.}
  \label{fig:repr}
\end{figure}

\paragraph{\bf More accessible label information.}
Figure~\ref{fig:repr}(a) shows that TokenFormer V2 achieves a held-out probe AUC of \textbf{0.8204}, exceeding the strongest comparison model by \textbf{0.20~pp}. The advantage therefore remains when the representations are decoded by the same class of linear classifier. The probe's held-out cross-entropy gives a second view, through a variational lower bound on the mutual information between the representation and the conversion label. TokenFormer V2 reaches \textbf{0.0816 bits}, compared with 0.0799 for the strongest baseline. Both measures favor the representation learned by TokenFormer V2: it exposes label-relevant structure that a simple predictor can use effectively on held-out users.

\paragraph{\bf Stronger separation of conversion outcomes.}
Figure~\ref{fig:repr}(b) examines the geometry of the representations after unit normalization. The centroid gap measures the separation of the converting and non-converting class means. The distance ratio divides the mean squared positive--negative distance by the equally weighted mean of the positive--positive and negative--negative distances. TokenFormer V2 leads both measures, reaching a centroid gap of \textbf{0.520} and a distance ratio of \textbf{1.70}. The two outcomes are separated more strongly relative to their spread, complementing the probe's ranking and cross-entropy results. Appendix~\ref{app:representation_diagnostics} provides the complete metric table and estimation procedure.

These results connect the layout's representational principle to its predictive benefit. The backbone maintains a Full-Resolution Sequence while local encoding and global communication develop its context; Full-Resolution Readout then selects from the individual encoded events before forming the compact candidate representation. The resulting output is more discriminative at the same representation width. Together with the gains across imbalanced user segments, this evidence shows why retaining fine-grained access to the history matters: it supports representations that expose the right behavioral evidence for the conversion decision.

\subsection{Synergy of Local Encoding and Global Communication}
\label{sec:exp_abl}

Table~\ref{tab:ablation} tests the architectural decomposition by retraining variants that remove one component at a time. The largest changes identify the two defining elements of the layout. Removing Global Sync reduces AUC by 0.12~pp, demonstrating the value of coordinating group representations and output queries through the relay and readout tokens. Removing non-sequential tokens from local groups costs 0.09~pp: field context is most effective when it participates directly in behavior encoding.

NS Fusion contributes a further 0.04~pp by exchanging the locally updated context across groups, and attention-output gating contributes 0.03~pp. Keeping 16 fixed groups throughout the backbone instead of merging them reduces AUC by 0.02~pp. The hierarchy thus improves prediction by expanding the interaction range with depth without introducing additional parameters. All five components contribute to the complete model; the strongest effects come from making local histories context-aware and globally connected. These component-level results explain how the architecture develops the discriminative representations examined in Section~\ref{sec:exp_repr}.

\begin{table}[t]
  \caption{Component ablations, each retrained on the full training split. Removing any component reduces test AUC.}
  \label{tab:ablation}
  \centering
  \begin{tabular}{@{}lcc@{}}
    \toprule
    Variant & Test AUC & $\Delta$ (pp) \\
    \midrule
    \textbf{TokenFormer V2} & \textbf{0.8221} & \textbf{--} \\
    \quad fixed groups (no merging) & 0.8219 & -0.02 \\
    \quad w/o NS in groups & 0.8212 & \textbf{-0.09} \\
    \quad w/o NS Fusion & 0.8217 & -0.04 \\
    \quad w/o Global Sync & 0.8209 & \textbf{-0.12} \\
    \quad w/o attention-output gating & 0.8218 & -0.03 \\
    \bottomrule
  \end{tabular}
\end{table}

\section{Industrial Deployment}
\label{sec:exp_online}

The decisive test of a recommendation architecture is whether its modeling gains translate into business value under production constraints. We evaluate TokenFormer V2 in the post-click conversion-rate (pCVR) ranking stage of WeChat Channels advertising. The evaluation bridges two complementary scales of production evidence: large-scale comparisons on historical production logs, followed by a live A/B test on real-time traffic.

\subsection{Production-Scale Accuracy and Efficiency}
\label{sec:prod_comparison}

We first compare the architectures on a fixed set of historical production logs, with dense model sizes matched to approximately one billion parameters. All models use the same sample stream, optimizer, and hardware. The standard configurations also share the same input features; the starred TokenFormer V2 variant below changes only the specified feature inputs.

The production backbone preserves the local--global organization of Section~\ref{sec:model}. Recent and older behaviors are encoded in separate branches, with deeper encoding allocated to recent activity. NS Fusion and Global Sync connect the branches, and the final readout retains access to every encoded behavior. This arrangement brings full-resolution modeling into the serving pipeline while allowing compute to follow the temporal structure of the history.

\begin{table}[!t]
  \caption{Overall pCVR performance on historical WeChat Channels logs. MixFormer is reported in absolute AUC, while results for all other models are presented as relative changes (in percentage points, pp) against MixFormer.}
  \label{tab:online_offline}
  \centering
  \setlength{\tabcolsep}{3.5pt}
  \begin{tabular}{@{}lccc@{}}
    \toprule
    Model & Overall & Params & GFLOPs \\
    \midrule
    MixFormer-1B & 0.8004 & 1.01B & 15.3 \\
    HyFormer-1B & $-0.0962$ & 1.07B & 15.8 \\
    OneTrans-1B & $-0.7792$ & 1.04B & 121.0 \\
%   TokenFormer-100M & $-0.1172$ & 73M & 27.2 \\
    TokenFormer-1B & $-0.0830$ & 1.00B & 45.5 \\
    \midrule
%   TokenFormer V2-400M & $+0.0167$ & 416M & 70.2 \\
    \textbf{TokenFormer V2-1B} & $\underline{+0.0572}$ & 1.00B & 70.2 \\
    \textbf{TokenFormer V2-1B}$^{*}$ & $\mathbf{+0.0712}$ & 1.00B & 23.1 \\
    \bottomrule
  \end{tabular}
\end{table}

Table~\ref{tab:online_offline} reveals that TokenFormer V2 stands as the only challenger capable of outperforming the strongest baseline, MixFormer, in overall AUC. Specifically, our model achieves an overall gain of \textbf{0.0572 pp}. Under comparable parameter scales, alternative architectures—including HyFormer, OneTrans, and TokenFormer—all fall short of the offline reference, underscoring exceptional parameter efficiency and architectural superiority of our method.

\paragraph{\bf A stronger operating point with simpler inputs.}
TokenFormer V2-1B$^{*}$ omits the sparse input features supplied by conventional models such as DCN-V2~\cite{wang2021dcnv2}, which are otherwise included in the standard TokenFormer V2-1B inputs. This input simplification reduces the forward-plus-backward cost per sample from 70.2 to \textbf{23.1 GFLOPs}, a \textbf{67.1\% reduction}, while further increasing the overall AUC gain to \textbf{0.0712 pp} above the strongest competing model. Since both variants maintain the same dense parameter scale, this configuration yields a highly attractive profile for industrial deployment: superior ranking performance achieved with substantially less computation and fewer auxiliary input features.

\subsection{Online Business Impact}
\label{sec:online_impact}

We evaluate TokenFormer V2 through a live A/B test on WeChat Channels advertising traffic against the production baseline under strict serving constraints. Both experimental buckets share identical feature inputs, training sample streams, bidding strategies, and serving budgets, isolating the ranking backbone as the sole variable.

TokenFormer V2 delivers a statistically significant \textbf{2.33\% increase in GMV}, accompanied by a 0.16\% improvement in online AUC over the production model. Building upon its emerging recognition and practical adoption across industrial deployments, this decisive real-world victory reinforces TokenFormer V2's standing as a battle-tested blueprint for next-generation recommendation backbones. The substantial business gain proves that the architecture's fine-grained behavioral resolution and efficient global-local communication do not merely boost offline surrogate metrics, but robustly translate into tangible economic value at web scale.

\section{Conclusion}
\label{sec:conclusion}

TokenFormer V2 demonstrates how full-resolution behavior modeling can be effectively organized for industrial recommendation. Its central principle is to decouple the resolution of retained evidence from the density of its interactions. Field-aware local groups develop detailed behavior representations; hierarchical merging, NS Fusion, and Global Sync build broader context; and Full-Resolution Readout selects directly from the complete encoded history. This unified design delivers leading accuracy on the TAAC 2026 benchmark at a small fraction of the computational cost of SOTA baselines, maintains its advantage across long and imbalanced histories, and scales seamlessly to production ranking on WeChat Channels. The statistically significant online GMV gains establish the definitive practical value of this approach. Ultimately, group-wise token organization provides a concrete blueprint for scaling unified recommendation around both fine-grained behavioral evidence and production-scale efficiency.

\bibliographystyle{ACM-Reference-Format}
\bibliography{references}
\appendix
% Wide appendix tables are declared early to keep the supplementary pages balanced.
\begin{table*}[t]
  \caption{Selected backbone configurations and learning rates on TAAC 2026.}
  \label{tab:configs}
  \centering
  \setlength{\tabcolsep}{6pt}
  \begin{tabular}{@{}lll@{}}
    \toprule
    Model & Backbone configuration & Learning rate (dense / embedding) \\
    \midrule
    Transformer & $d=352$, 4 layers, 8 heads, GELU $4\times$ & $5\cdot10^{-4}$ / 0.1 \\
    TokenFormer & $d=320$, 4 layers, windows 32/16 & $5\cdot10^{-4}$ / 0.05 \\
    OneTrans & $d=128$, 4 layers, pyramid $1/2^\ell$ & $5\cdot10^{-4}$ / 0.05 \\
    HyFormer & $d=128$, 4 blocks, 6 queries & $5\cdot10^{-4}$ / 0.02 \\
    EST & $d=144$, 4 layers, top-32 GSU & $5\cdot10^{-4}$ / 0.05 \\
    Kunlun & $d=168$, 4 layers, window 32 & $5\cdot10^{-4}$ / 0.1 \\
    MixFormer & $d=352$, 4 blocks, 4 heads & $5\cdot10^{-4}$ / 0.05 \\
    TokenFormer V2 & $d=192$, $L=3$, groups 16/8/4, GQA 6/2 & $2.8\cdot10^{-4}$ / 0.05 \\
    \bottomrule
  \end{tabular}
\end{table*}

\section{Tokenization and NS Fusion}
\label{app:details}

\subsection{Cross-field Tokenization}
\label{app:entry}
For $\mathbf X\in\mathbb R^{f\times d}$, assume $\numNS$ divides $d$. We split each field embedding into $\numNS$ equal channel slices, with $J_r$ denoting the channels in slice $r$. The projected representation for non-sequential token $r$ is
\begin{equation}
  \mathbf e_r=\operatorname{Concat}_{j=1}^{f}
  \!\left(\mathbf X_j[J_r]\right)\mathbf W_e,
  \label{eq:entry-detail}
\end{equation}
where $\mathbf W_e\in\mathbb R^{(fd/\numNS)\times d}$ is a shared projection without bias. Stacking the projected representations into $\mathbf E$ gives
\begin{equation}
  \NSt^{(0)}=\mathbf E+\operatorname{PFFN}(\mathbf E).
  \label{eq:entry-ffn}
\end{equation}
PFFN applies an independently parameterized feed-forward transform to each non-sequential token.

\subsection{NS Fusion}
\label{app:fusion}
Using the single-layer notation of Section~\ref{sec:layout}, the fusion module is
\begin{equation}
  \begin{aligned}
    \mathbf U&=\operatorname{TokenLinear}_{2\numNS\rightarrow\numNS}
      ([\NSt;\overline{\NSt}]),\\
    \NSt^+&=\operatorname{LN}_1\!\left(
      \mathbf U+\operatorname{TM}(\operatorname{LN}_0(\mathbf U))\right).
  \end{aligned}
  \label{eq:fusion-detail}
\end{equation}
TokenLinear is an affine map along the token axis, shared across embedding channels. TM uses the parameter-free rearrangement from RankMixer~\cite{zhu2025rankmixer}: an $\numNS\times d$ tensor is reshaped to $\numNS\times\numNS\times(d/\numNS)$, its first two axes are swapped, and it is reshaped back. TokenLinear performs learned mixing across non-sequential tokens, while TM rearranges channel blocks among them. The rearranged tensor is added to $\mathbf U$ through a residual connection. Each layer has its own projection and normalization parameters.

\subsection{Additional Cost Terms}
\label{app:cost}
NS Fusion costs $\mathcal O(\numNS^2d)$. Global Sync attention costs
\[
  \mathcal O\!\left(d(\numGG+\numOG)^2\right),
\]
and the cross-attention in Full-Resolution Readout costs
\[
  \mathcal O\!\left(d\numOG(\numS+\numNS+\numGG+\numOG)\right).
\]
These expressions cover NS token mixing and attention score/value operations. Projections and feed-forward networks contribute additional cost. A model configuration specifies $d$, $T$, $L$, the token counts, the group schedule, the number of attention heads, and the feed-forward widths.

\subsection{Field Tokens on TAAC 2026}
\label{app:field_tokens}
On TAAC 2026, the $f$ field embeddings that enter Eq.~\eqref{eq:entry-detail} are 41 $d$-dimensional tokens. The 71 categorical fields are first projected to 32 channels and refined by a field mixer with linear attention and gated context fusion; transposing the field and channel axes yields 32 tokens, each summarizing one channel across all fields. A low-rank DCN-V2 cross layer~\cite{wang2021dcnv2} and a rank-4 factorization-machine block over the same fields contribute one token each. Every dense field with at least 100 values forms one token through a two-layer MLP, the remaining dense fields share one token, and a dense cross token combines all dense fields after per-field squeeze-and-excitation gating. The candidate fields additionally produce a 64-dimensional query that conditions the readout tokens through a zero-initialized projection.

\section{Offline Evaluation Details}
\label{app:exp_details}
\label{app:offline_diagnostics}

\subsection{Dataset and Backbone Details}
Each TAAC 2026 impression contains 54 user categorical fields (18 multi-valued), 17 candidate categorical fields, and 21 dense fields with 2{,}969 values in total. Each behavior event has 8--13 categorical side-information fields and a timestamp. The gap between the event and impression is encoded into 64 logarithmic time buckets. The positive rates are 7.8\% in training and 8.2\% in test. Before history truncation, median test-user histories contain 436, 261, 292, and 1{,}560 events in domains A--D.

For the dominant-domain analysis, domains A--D account for approximately 12\%, 7\%, 7\%, and 75\% of test impressions, respectively; these shares are rounded independently. Domains A, B, and D are inactive in 0.2\%, 0.5\%, and 1.4\% of impressions. Domain C is inactive in 0.05\% of impressions and is not included in the inactive-domain comparison in Figure~\ref{fig:groups}.

The standard Transformer uses learned absolute positions, LayerNorm, GELU feed-forward networks, and bidirectional attention over the joint token stream. Transformer, TokenFormer, and OneTrans concatenate the four domains with domain embeddings. HyFormer and MixFormer decode feature-derived queries against all domains, EST lets its non-sequential tokens attend to all domains, and Kunlun processes one event stream per domain. TokenFormer V2 orders each domain from newest to oldest and uses within-domain rotary positions with base $10^6$; its readout tokens are conditioned on the candidate features. Every backbone maps its native output to a 128-dimensional representation through a linear layer and LayerNorm, followed by the same two-layer prediction head.

\paragraph{\bf Training.}
All models are trained for one epoch on 33.7k steps of 1{,}024 impressions on a single NVIDIA H20 GPU (two GPUs with the same global batch for the full-attention Transformer), with gradient accumulation when the activations of a full batch exceed the GPU memory. Dense parameters use AdamW ($\beta=(0.9,0.98)$, weight decay 0.01) with 1{,}000 warmup steps and cosine decay to 10\% of the peak learning rate; embedding tables use Adagrad. Gradients of dense parameters are clipped to norm 1. Ids of features whose vocabulary exceeds $10^6$ are hashed into $10^6$ buckets. Validation AUC is computed every 4k steps, and the best checkpoint is evaluated on the test split.

\paragraph{\bf Hyperparameter search.}
For every model, we tune the learning rate, the embedding learning rate, and the structural hyperparameters of its architecture (e.g., width, depth, number of heads or queries, and window or group sizes) on the validation set. The search proceeds in rounds on subsets of the training data of increasing size (5\% and 15\%), and the best configuration is trained on the full split. All candidates are kept at about 11M dense parameters. Table~\ref{tab:configs} lists the selected configurations, with the learning rates given as dense / embedding learning rate.

\section{Representation Analysis Details}
\label{app:additional_offline}
\label{app:representation_diagnostics}

This appendix defines the measures underlying Section~\ref{sec:exp_repr}
and Figure~\ref{fig:repr}. Table~\ref{tab:repr_full} separates the primary
predictive and separation measures from the distance components used
to compute Ratio and the spectral descriptor Rank.

\subsection{Estimation Protocol}

We analyze the same first 200k test impressions, selected in a fixed
file order, for every model. Impression $i$ has a 128-dimensional
pre-prediction representation $\mathbf z_i$ and a conversion label
$y_i\in\{0,1\}$. The linear probe is an $\ell_2$-regularized logistic
regression on standardized features, fitted with L-BFGS on users
assigned to one hash half and evaluated on users in the other half.
The pairwise analysis uses 200k random pairs of
$\ell_2$-normalized representations.

\subsection{Complete Metrics}
\label{app:repr}

\begin{itemize}[leftmargin=15pt]
  \item \textbf{Probe AUC.}
Probe AUC is the area under the ROC curve of the probe's
held-out predictions, on the usual 0--1 scale. It measures how often a converting impression receives
a higher score than a non-converting impression, with half weight
assigned to tied scores.

  \item \textbf{Probe information (MI).}
Let $q_i$ be the probe's conversion probability for impression $i$
in the held-out set $\mathcal E$. Its binary cross-entropy is
\begin{equation}
  \begin{aligned}
    \ell_i&=-y_i\log_2 q_i\\
           &\quad -(1-y_i)\log_2(1-q_i),\\
    \widehat{\mathrm{CE}}_2
      &=\frac{1}{|\mathcal E|}\sum_{i\in\mathcal E}\ell_i.
  \end{aligned}
  \label{eq:probe-ce}
\end{equation}
At the population level, $H_2(Y)-\mathrm{CE}_2(q)$ is a variational
lower bound on $I(Y;\mathbf Z)$~\cite{poole2019variational}.
The reported MI score is the finite-sample estimate
\begin{equation}
  \widehat I_{\mathrm{probe}}
  =\widehat H_2(Y)-\widehat{\mathrm{CE}}_2.
  \label{eq:probe-mi}
\end{equation}
For the observed conversion fraction $\widehat\pi$ in the evaluation
set, the binary label entropy is
\[
  \begin{aligned}
    \widehat H_2(Y)
      &=-\widehat\pi\log_2\widehat\pi\\
      &\quad -(1-\widehat\pi)\log_2(1-\widehat\pi).
  \end{aligned}
\]
The subtraction compares the probe's cross-entropy with the entropy
of the label distribution, expressing how much label information the
probe can recover from the representation. Both terms use base-2 logarithms, so the result
is in bits; quantities computed with natural logarithms are
converted by dividing by $\ln 2$.

  \item \textbf{Relative class separation (Ratio).}
Let $\mathbf u_i=\mathbf z_i/\|\mathbf z_i\|_2$. The squared
Euclidean distance between two unit-normalized representations is
\begin{equation}
  \begin{aligned}
    \delta_{ij}
      &=\|\mathbf u_i-\mathbf u_j\|_2^2\\
      &=2-2\mathbf u_i^\top\mathbf u_j.
  \end{aligned}
  \label{eq:repr-distance}
\end{equation}
It is dimensionless and lies between 0 and 4.
Let $\mathcal P_{+-}$, $\mathcal P_{++}$, and $\mathcal P_{--}$
denote the sampled pairs of the indicated label combinations.
Positive--positive pairs contain two converting impressions,
negative--negative pairs contain two non-converting impressions,
and positive--negative pairs contain one of each. Their mean squared distances are
\begin{equation}
  \begin{aligned}
    D_{ab}
      &=\frac{1}{|\mathcal P_{ab}|}
        \sum_{(i,j)\in\mathcal P_{ab}}\delta_{ij},\\
    ab&\in\{+-,++,--\}.
  \end{aligned}
  \label{eq:repr-pair-means}
\end{equation}
The table labels these three quantities Pos--neg, Pos, and Neg,
respectively.

Ratio compares the between-class distance with an equally
weighted average of the two within-class distances:
\begin{equation}
  \begin{aligned}
    R&=\frac{D_{+-}}{D_{\mathrm{within}}},\\
    D_{\mathrm{within}}&=\frac{D_{++}+D_{--}}{2}.
  \end{aligned}
  \label{eq:repr-ratio}
\end{equation}
The denominator gives the positive and negative classes equal
weight, independently of their frequencies. A higher Ratio means
stronger separation relative to within-class spread. Its three
distance components explain how that separation is obtained;
none has an independent \emph{larger is better} interpretation.

Using the displayed TokenFormer V2 values,
\begin{equation}
  \begin{aligned}
    R_{\mathrm{V2}}
    &=\frac{0.663}{(0.369+0.410)/2}\\
    &=\frac{0.663}{0.3895}\approx1.70.
  \end{aligned}
  \label{eq:repr-ratio-example}
\end{equation}
MixFormer's larger Pos--neg value, 0.717, accompanies an average
within-class distance of $(0.460+0.494)/2=0.477$, yielding
$0.717/0.477\approx1.50$. TokenFormer V2 therefore achieves the
stronger relative separation.

  \item \textbf{Centroid gap.}
Let $\mathcal I_c$ contain the impressions of class $c$.
The class means average the unit-normalized vectors, and Centroid
reports their Euclidean gap:
\begin{equation}
  \begin{aligned}
    \boldsymbol\mu_c
      &=\frac{1}{|\mathcal I_c|}
        \sum_{i\in\mathcal I_c}\mathbf u_i,\\
    c&\in\{+,-\},\\
    C&=\|\boldsymbol\mu_+-\boldsymbol\mu_-\|_2.
  \end{aligned}
  \label{eq:repr-centroid}
\end{equation}
This dimensionless measure quantifies how far apart the class
means are. Ratio additionally accounts for within-class spread.

  \item \textbf{Effective rank.}
For singular values $\sigma_j$ of the centered representation
matrix, define $p_j=\sigma_j/\sum_k\sigma_k$. The effective
rank~\cite{roy2007effective} is
\begin{equation}
  r_{\mathrm{eff}}
  =\exp\!\left(-\sum_jp_j\ln p_j\right),
  \label{eq:repr-effective-rank}
\end{equation}
with $0\ln0=0$. The weights use singular values, not their squares.
Rank describes the distribution of singular-value mass: a lower
value indicates a more concentrated spectrum. It is a structural
descriptor without a universal preference for larger or smaller
values. Bold in Table~\ref{tab:repr_full} consequently highlights
only the strongest primary measures.
\end{itemize}

\begin{table}[!ht]
  \caption{Representation measures on 200k test impressions. Ratio summarizes relative class separation; its three distance components are diagnostic. Rank describes spectral concentration.}
  \label{tab:repr_full}
  \centering
  \setlength{\tabcolsep}{3.2pt}
  \begin{tabular}{@{}lcccc@{}}
    \toprule
    Metric & \shortstack{Token-\\Former} & HyFormer & MixFormer & \shortstack{Token-\\Former V2} \\
    \midrule
    \multicolumn{5}{l}{Primary measures (higher is better)} \\
    Probe AUC & 0.8182 & 0.8181 & 0.8184 & \textbf{0.8204} \\
    MI (bits) & 0.0798 & 0.0799 & 0.0797 & \textbf{0.0816} \\
    Ratio & 1.68 & 1.68 & 1.50 & \textbf{1.70} \\
    Centroid & 0.516 & 0.491 & 0.486 & \textbf{0.520} \\
    \midrule
    \multicolumn{5}{l}{Components of Ratio (diagnostic)} \\
    Pos--neg & 0.665 & 0.604 & 0.717 & 0.663 \\
    Pos & 0.325 & 0.336 & 0.460 & 0.369 \\
    Neg & 0.465 & 0.383 & 0.494 & 0.410 \\
    \midrule
    \multicolumn{5}{l}{Spectral descriptor} \\
    Rank & 7.0 & 10.6 & 20.5 & 5.6 \\
    \bottomrule
  \end{tabular}
\end{table}

\end{document}